# Why Johnny Can't Store Passwords Securely?

A Usability Evaluation of Bouncycastle Password Hashing

Chamila Wijayarathna
University of New South Wales
Canberra, ACT
c.diwelwattagamage@student.unsw.edu.au

Nalin A. G. Arachchilage
University of New South Wales
Canberra, ACT
nalin.asanka@adfa.edu.au

## ABSTRACT

Lack of usability of security Application Programming Interfaces (APIs) is one of the main reasons for mistakes that programmers make that result in security vulnerabilities in software applications they develop. Especially, APIs that provide cryptographic functionalities such as password hashing are sometimes too complex for programmers to learn and use. To improve the usability of these APIs to make them easy to learn and use, it is important to identify the usability issues exist on those APIs that make those harder to learn and use. In this work, we evaluated the usability of SCrypt password hashing functionality of Bouncycastle API to identify usability issues in it that persuade programmers to make mistakes while developing applications that would result in security vulnerabilities. We conducted a study with 10 programmers where each of them spent around 2 hours for the study and attempted to develop a secure password storage solution using Bouncycastle API. From data we collected, we identified 63 usability issues that exist in the SCrypt implementation of Bouncycastle API. Results of our study provided useful insights about how security/cryptographic APIs should be designed, developed and improved to provide a better experience for programmers who use them. Furthermore, we expect that this work will provide a guidance on how to conduct usability evaluations for security APIs to identify usability issues exist in them.

## 1 INTRODUCTION

In 2017 December, a group of researchers discovered a database on dark web that contains 1.4 billion user-names and passwords in clear text [7]. We are hearing similar incidents of password leaks every now and then [7, 9]. Even though so many advanced technologies have been introduced to secure passwords from hackers, still hackers have been successful in conducting attacks and stealing user passwords. One of the main reasons for this is developers, who develop the applications and responsible for the security of passwords, are making mistakes while developing applications, which results in making password stores accessible to hackers. Due to such issues caused by developers, researchers have called the developer as the enemy of cyber security [15].

Storing user login data and authenticating users is one of the most common tasks carried out by software developers [10]. At the same time, this task is prone to security issues due to the high complexity of the technologies and concepts involved in this process [6]. Most developers have been identified to store passwords and other login details insecurely due to their lack of expertise in these security related concepts and technologies [10].

One of the possible solutions to solve this problem is to make the libraries and Application Programming Interfaces (APIs), which provide such security related functionalitties, easy to use for programmers by improving their usability [15]. However, researchers have been often blaming such APIs, which we call as security APIs or crypto APIs, as less usable and difficult to use correctly by programmers [6].

Therefore, in this study, we are evaluating the usability of the password hashing functionality provided in one of the most popular cryptographic APIs that provides cryptographic functionalities such as password hashing, Bouncycastle API . Bouncycastle is an open source library that implements cryptographic functionalities and exposes them to programmers to use through an API. Usability issues exist in such security APIs force non security expert programmers to use the APIs incorrectly and result in introducing security vulnerabilities to applications they develop. To improve the usability of APIs, it is important to understand what programmers expect from APIs and why API implementations have failed to meet programmer expectations. In this work, by observing programmers who use Bouncycastle API to implement a secure password storage, we attempt to identify usability issues exist in the API that persuade programmers to insecurely store passwords. Following is the research question that we are trying to answer in this work.

- What are the usability issues exist in password hashing functionality of Bouncycastle API that make programmers incorrectly use the API and how those issues affect security of the applications that programmers develop?

To answer this question, we conducted a qualitative experimental study with 10 participants where each participant spent about 2 hours for the study. In this experiment, we used cognitive dimensions questionnaire based methodology [14] and think aloud method [13] to identify usability issues of the API that programmers experienced. From the data we gathered, we identified useful insights about usability issues exist in Bouncycastle API. Following are some of the key usability concerns of the API that we identified.

- Difficulty in identifying correct parameters to use in API method invocation.
- Difficulty in using the API correctly without a previous knowledge on security concepts such as salt.
- Poor documentation.
- Lack of examples.

We believe the knowledge obtained in this study will help to develop more usable security APIs in the future and will guide developers to improve the usability of existing security APIs.

The paper is organized as follows. Section 2 reviews previous related research. Section 3 describes the experiment methodology and section 4 presents the results of the study. Then we discuss the impacts of usability issues we identified in this study to the security of applications before concluding the paper.

## 2 RELATED WORK

The relationship between the usability of security APIs and security of end user applications that use those security APIs has become a topic of high interest among researchers recently [1, 2, 5, 6, 8, 14] . There have been several studies that discuss and investigate this relationship.

By emphasizing that programmers are not security experts, Wurster and van Oorschot argue that improving usability of tools and APIs that programers use is important to minimize the mistakes they do while developing applications [15]. Acar et al. [1] also highlight the importance of the usability of security APIs by pointing that programmers who make use of security APIs are not experts of security. Mindermann [8] argues that security of an application will be far better if the libraries used to develop that application are more usable. He stresses the importance of applying usability research for security APIs to deliver more usable security APIs.

Even though the importance of the usability of security APIs has been identified and discussed [1, 2, 8, 15], not much work has been done to evaluate the usability of security APIs. Gorski and Iacono [5] presented 11 characteristics that need to be considered when evaluating the usability of security APIs. Green and Smith [6] introduced 10 rules for developing usable security APIs. By considering these 2 sets of guidelines and by referring to previous work done on usability evaluation of general APIs, Wijayarathna et al. [14] presented a cognitive dimensions framework, which consists of 15 dimensions to be used in the usability evaluation of security APIs. The only current work on empirically evaluating usability of security APIs are done by Acar et al. [1] where they evaluated and compared usability of 5 cryptographic APIs for python.

There are several studies that investigate the programmer behavior related to password storage. By doing an empirical investigation on password storage behaviours of programmers, Naiakshina et al.[10] revealed some interesting insights about the password storage behaviour of developers such as security knowledge does not guarantee secure password storage. In a study conducted by Acar et al. [2] to evaluate the effectiveness of recruiting participants through Github for security developer studies, they have used a password storage task and they have observed that some participants stored passwords as plain text. Authors have observed that this insecure behaviour does not depend on their self reported security knowledge.

However, so far there has been no study that investigates usability issues of cryptographic APIs that persuade programmers to make mistakes while working with user passwords. Our work attempts to fill this gap by studying programmers who use Bouncycastle crypto API to secure passwords and by identifying usability issues of the API.

## 3 METHODOLOGY

The study was designed to identify issues that programmers face as well as to observe how those issues affect them while they are using Bouncycastle API for implementing secure password storage solutions. This study was approved by the Human Research Ethic Committee of our university.

Conducting a user study is a widely known method for identifying usability issues of APIs [1, 14]. To conduct a user study, first we had to design a programming task for participants to follow. Even though Boucycastle provides numerous password hashing algorithm implementations, it is not practical to evaluate all these functionalities in a single study. On the other hand, Naiakshina et al. [10] observed that programmers are not good at selecting secure algorithms to use in password hashing.Therefore, we decided to inform participants on which algorithm to use and we decided to use SCrypt algorithm [11] as it was the most secure algorithm implementation available in Bouncycastle at the time we conducted the experiment. However, we observed that results we obtained are relevant to other hashing algorithms of Bouncycastle as well, as these different implementations have lots of similarities in the design.

In the task, we provided participants with a simple web application that includes functionalities for registering users and login users. A user can register for the application by providing a username and a password. Once the registration is completed, user can login to the web application with the username and the password. In the web application that we provided for the participants, we stored passwords as plain text and we asked participants to secure passwords using SCrypt implementation of the Bouncycastle API.

We recruited programmers with Java experience from Github to participate in the study. We used Github to recruit participants rather than recruiting participants from university or local software development firms, to get a more diverse sample of programmers. This is a widely accepted and used method among researchers to recruit participants to developer studies [1, 2]. Furthermore, recruiting participants from Github helps to get participants with more experience in software development, which improves the ecological validity of the study. We extracted publicly available email addresses of Java developers with significant contributions to Java projects and sent emails inviting them to participate in our study. We offered them with a $15 Amazon gift voucher as a token of appreciation for the participation. In the invitation email, we included a link to sign up for the study. Furthermore, we informed them that participation is voluntary and participants can withdraw from the study at any time. Once people signed up, we filtered out those who did not have any software development experience since our target sample for the study was software developers. Furthermore, we filtered out participants with no experience in using Java because if a participant faces issues with programming language while completing the task, we may not be able to clearly identify the usability issues of the API s/he had come up with. Sign up form required participants to enter their name and email address, which were required to send study material to them. However, such personally identifiable information of the participants were removed from the final data set which we used for the analysis. A total of 10 participants participated in the study.

Once each participant signed up by completing the sign up form and consented to participate in the study, we sent them details of the programming task to do and code artefacts to use. Participants completed the task remotely on their own computers and we suggested them to complete the task in a time comfortable to them. We requested them to think-aloud [13] and record their screens with voice (so thinkaloud results will be recorded) while completing the task. By collecting participant think-aloud results, we expected to identify issues they experience by observing their thought process. Once a participant completed the task, s/he was asked to send his/her source code with the video recording to us through email. Participants spent an average of 44.32 minutes in completing the task. Then each participant had to complete the cognitive dimensions questionnaire [14] which we shared with them via Google forms.

Once we finished the data collection, analysis was done manually by one analyst. We used manual analysis since our data set was small[3]. Questionnaire answers were analysed prior to analysing videos and identified usability issues that exist in Bouncycastle API from questionnaire responses. After analysing questionnaire answers, recordings were analysed to identify the usability issues that each participant encountered. Then, those were analysed to identify how usability issues that were identified, affected the participants for securely completing the task. Special attention was given to decisions made by participants that caused to reduce the security of programmes they developed.

## 4 STUDY RESULTS

From the analysis, we could identify a total of 63 issues in the SCrypt implementation of Bouncycastle. Each participant had reported an average of approximately 15 usability issues. From here onward, we are looking at issues that were experienced by most number of participants and those that had an effect in the security of the password storages developed by the participants. Furthermore, we have logically divided the identified issues into several categories for the ease of presentation and discussion.

For the ease of presentation, we labelled participants with labels P1, P2,.., P10. They will be referred with this label from here onward. Statements made by participants that are presented in this section were not corrected for any grammatical errors and are presented as those were stated.

### 4.1 Selecting method parameters for SCrypt generate method invocation

Bouncycastle API exposes its SCrypt hashing functionality with a static method "*generate()*" of *SCrypt* class. This method takes 6 parameters, which are password as a byte array, salt value as a byte array, CPU/Memory cost parameter, blocksize in use for underlying hash, parallelization parameter and the length of the key to generate . The strength of the password storage depends on the values that programmer uses for these parameters. If the programmer uses weak values for these parameters, security of the password storage will be weak and it will be vulnerable to attacks.

Most of the participants of the study identified that using correct values for these parameters is important to ensure security. 7/10 participants mentioned this in their response to the questionnaire and mentioned that if they used incorrect values, it would negatively impact the security of their solution. P7 mentioned that "*Application would be secure if I used the appropriate values for the parameters N, P and R. For example if N is very small I realized that the processing is faster but I suppose it would be easier to hack the password*". However, understanding these parameters and selecting correct values for these parameters were not that easy and some of our participants blamed the API for not providing sufficient help for this.

Some participants (P1, P2, P4, P5, P7, P8, P9) found it difficult to understand what these parameters mean and found it difficult to interpret them. Furthermore, participants mentioned that parameters are too low level and it was difficult to find required details about those parameters that are required to use the API. Participants reported this in their questionnaire responses as well as we could observe this in their think aloud results. While completing the task, we could observe these thoughts from P2 - "*What is parallelization? What does parallelization does? I'm thinking what does parallelization use for. It would make sense if I would read some stuff about SCrypt in the web. For now, let's try to*

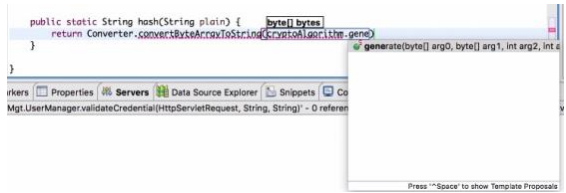

(a) Suggestions shown in Integrated Development Environment for method parameters

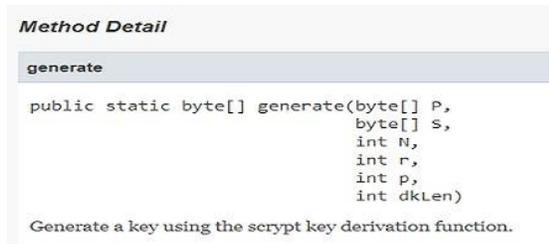

(b) Method parameter names used in Bouncycastle SCrypt source code

![Java API documentation of SCrypt.generate() method]

(c) Java API documentation of SCrypt.generate() method

Figure 1: Examples of non-explanatory variable names observed by participants

guess what is happening. Not a good thing to guess when you are doing security". However, in a good API, programmers are supposed to understand what a class/method/variable does just by looking at its name, which has not been the case here.

API Java docs or the Integrated Development Environment(IDE)'s suggestions also did not gave any help for programmers to understand these parameters as they show single letter non-explanatory variables. API source code which can be accessed through internet also uses non-explanatory variable names. Figure 1 shows some of the places where participants observed non-explanatory variable names of method parameters that made it difficult for them to interpret meaning and use of those parameters. P7 mentioned in his response to the questionnaire that "*It was difficult to find the details of the API while you are using it. programming IDE suggests using the API with unclear input argument names (arg1, arg2, etc.)*".

Most participants mentioned that selecting values for these parameters as the hardest part of using the API. Even participants who could understand the use of parameters also found it difficult to identify proper values to use as parameter values. Participants urged that API should provide some guidance to select values for parameters when calling the *SCrypt.generate()* function. We observed their frustration in their think aloud results as well as in post experiment survey feedback. P8 while trying to identify proper values for parameters said that "*CPU memory cost parameter, I have no idea how to choose this number, [..], Block size must be greater than 1, if 100 is okay or 1000, I don't know*". Furthermore, participants suggested that if the API and its documentation can provide information about the relationship between the values used for parameters and strength of produced hash as well as relationship between the values used for parameters and performance of the algorithm, it would make the life of the programmer less difficult.

Participants also hinted some solutions that would reduce the difficulties programmers face while choosing parameter values. Some participants (P4, P5, P7, P9) suggested that SCrypt class should provide an overloaded generate() method which only takes password or password and salt as method parameters. For example, P7 suggested that "*Maybe add another method that masks the call to Scrypt's generate () method using default values for N, P, and R*". It might not be possible to provide functionalities in the exact way that participants have suggested here. However, this suggests that there is a clear mismatch between what programmers expect and what the API has provided.

Participants also suggested to use more high-level objects for method parameters that enforce constraints associated with each parameter, rather than using primitive Java *int* type. P10 explained this saying "*The method only uses primitive types for its arguments. This is pretty low level. Only the documentation gives a hint on what subset of integers are valid. Real objects as arguments could provide much better documentation and make their usage much more obvious. Some restrictions are given in the documentation as formulas, these restrictions could be coded in the argument types.*".

### 4.2 Using salt with *SCrypt.generate()* method

Salt is also a parameter of the *SCrypt.generate()* method, but it is different from cost parameters we discussed in previous sub sections and serves a different purpose. Usage of salt provides protection against dictionary attacks and pre-computed rainbow attacks. In study results, we observed that different ways programmers followed to use salt affected the security of password storage solutions they developed.

Even though most of the participants identified that values they select for cost parameters would affect the security of their solution, we observed that they did not feel the same about the salt value they used. Interestingly, most of the participants used a constant salt among all the users of their application. We observed that this was mainly a result of their lack of knowledge on the salt concept and their confidence about their knowledge which made them believe that they are doing the right thing. They did not seem to knew that they were doing something bad as it was not reflected in their thoughts. P8 created 2 users with same password and observed that both get the same hash, still did not identify it as a problem.

Interestingly, API did not give any guidance for the participants to avoid false uses such as using constant strings as the salt. It has been acknowledged that usable security APIs should be possible to use by programmers correctly and securely, even without a previous knowledge of security concepts. However, it was appeared that the API has assumed that programmers who will use the API would already be knowledgeable about relevant concepts and would use salt as expected by API developers. Some participants (P1, P4, P5, P8) also stressed that Bouncycastle API should give proper guidance about how to use the salt when using this functionality. Furthermore, participants suggested that the API should provide recommendations on the length of the salt to be used. P8 mentioned in his think aloud results that "*I know what the salt means, I can just put any random number there, well, we got to know if there should be any criteria for choosing that byte array*".

### 4.3 Issues observed about the documentation of Bouncycastle API

Documentation is an integral part of any API and it has a huge effect on the experience that programmers have with an API. In Bouncycastle API documentation also, there were few issues that participants experienced and suggested that could have been fixed to improve the usability of the API. We observed that participants were referring to unreliable third party sources and tutorials rather than referring to official documentation, which is considered more reliable[4]. We observed that participants were referring to Stackoverflow to get help about using the API and also for making security critical decisions such as selecting values for method parameters. Apparently, it was observed that most details about the parameter selection and recommendations for the parameters were only available in Stackoverflow.

After completing the task, some participants (P1, P4, P5, P7) mentioned that there was no enough documentation to refer. P4 mentioned that "*Not much documentation for this class [SCrypt] and very few examples in the internet for this API* ". Furthermore, we observed that participants found it difficult to find documentation and tutorials about the SCrypt implementation that they had to use. They blamed that there were no proper link from the Bouncycastle website to the API documentation and they said that information was difficult to find. P7 mentioned in his think aloud output that "*This looks simple to use, The problem is that I do not have the documentation or I can't find the documentation*".

Participants also mentioned that there was no enough examples on how to use the SCrypt functionality of the API. P5 mentioned in his response to the questionnaire that "*The biggest problem was finding an example of how to use the required part of the API. I could not find examples in the API site and had to refer to user groups on the web for detail.*"

We made another interesting observation when participants search for "Bouncycastle SCrypt" in Google , where results from third party sources such as Stackoverflow appeared before sources from Bouncycastle. This encouraged some participants to refer to solutions provided in Stackoverflow before looking at official documentation. It has been acknowledged previously that referring to third party sources such as Stackoverflow can negatively affect the security of programs that are developed [4] and what we observed with Google search results encouraged that negative behavior. We also observed that some of the Stackoverflow posts that were referred by participants even had compile errors in the available code, which made their life much harder. This observation suggests that it is important to list official sources in the top of Google results before other third party sources. It would be an interesting thing to explore whether there is a significant relationship between the order of Google search results and the security of applications developed by programmers.

### 4.4 Usage of byte array vs String to store passwords

Using Java char arrays to store passwords is recommended over using Strings to store passwords, because while using Strings, password will be stored in the memory for longer and cannot be deleted until garbage collector clears them. Because of this, cryptographic API methods such as SCrypt.generate() use char arrays and byte arrays as the variable type for passwords.

However, participants were not very pleased to use byte arrays as they preferred to use String over byte array. P5 suggested that "*The API used byte arrays which are similar to Strings. Again,a String overload would have been a nice addition.*". However, what programmers expected was not the most secure option and it would compromise the security. We identified this as a place where usability has been compromised to provide better security. It would be useful to explore whether these programmer expectations can be fulfilled without compromising the security.

Furthermore, we observed that some participants (P2, P4, P8, P10) used String to save passwords in their applications before using the hashing function and only converted it to byte array for the hashing step. Even though the API has taken measures to prevent programmers from using String, it has not been successful as these details have not been properly communicated to the programmer. This suggests that providing more explanation on security details with API documentation would help to prevent programmers from taking insecure decisions, which will enhance the security of the applications they develop.

### 4.5 Other observed issues

Bouncycastle provides password hashing functionality with SCrypt algorithm through the *generate()* static method of the *SCrypt* class. Even though, this method is static, API allowed programmers to instantiate SCrypt objects. Being able to instantiate the class and then the method being static confused some programmers. P7 described his confusion saying "*I have a confusion initially because I create an object of class Scrypt, but later I see that the method generate() was static. The Scrypt class should have private constructor.*"

## 5 DISCUSSION

While our main intention of this study was to identify usability issues of the Bouncycastle SCrypt implementation, we observed some interesting insights about the security usability relationship, specially in the API context.

There are several studies on how method parameters of API methods affect the experience of programmers [12]. However, not much emphasis is given to how it would affect the security of end user applications that use the API when such design decisions go on the bad way. Previous research has discussed that factors such as "error proneness" or "lack of documentation" can affect the security of end user applications that use a security API. Our results show that usability issues of more primitive level, such as issues related to abstraction level of the security API can also affect the security of end user applications that use the API. This emphasizes and highlights the importance of improving usability of security APIs to prevent security vulnerabilities in applications that make use of those APIs.

Our results also highlight the importance of designing security APIs such as Bouncycastle for programmers who are not security experts. We observed how programmers without proper knowledge of salt could misuse the API to develop a solution that is not secure. Previous research has highlighted a lot that developers who use security APIs such as Bouncycastle are not security experts and it is important to design APIs assuming that the programmer who would use this API is not a security expert [5, 6, 15].

Due to the usability issues experienced that made the SCrypt implementation of Bouncycatle difficult to use, P3 in his post task survey response mentioned that "*Sincerely, I'd just had used SHA-256. No need to reinvent the wheel.*". This suggests that if more secure options are provided with less usability, programmers would go for the less secure options. This is an important observation and highlights why improving usability of security APIs is so important. This also agrees with the findings of Naiakshina et al. [10] where they identified that most programmers are not competent enough to select password hashing algorithms to secure their password stores. In organizational scenarios, organizations should enforce programmers to use particular hashing algorithms with enough strengths to secure their applications rather that expecting programmers to select algorithms.

Even though we evaluate the SCrypt functionality of the Bouncycastle API, most issues we identified in this study are applicable to some of the other hashing implementations of Bouncycastle such as BCrypt.

## 6 CONCLUSION

In this study, we conducted a remote behavioural usability study with 10 software developers to identify usability issues that exist in Bouncycastle SCrypt implementation. Participants were asked to complete a simple programming task where they had to develop a secure password storage using Bouncycastle API. They had to think aloud and record their screens while completing the task and once they finished the task, they had to answer the cognitive dimensions questionnaire. Through the data we collected, we were able to identify usability issues that exist in the API.

We mainly observed that participants faced issues while using the API due to complexity of *SCrypt.generate()* method parameters and due to lack of knowledge in security concepts such as salt. We analysed and discussed how usability issues we identified in this study affected the experience of programmers and the security of applications they develop. We expect our results would help to improve usability of similar security APIs that will be developed to provide similar security and cryptographic functionalities.We also expect that this work will provide a guidance for researchers and software engineering community on how to conduct usability evaluations to identify usability issues exist in security APIs.